\documentclass[twocolumn]{aastex631}
\usepackage{natbib}
\usepackage{chemformula}
\usepackage[T1]{fontenc}
\usepackage{times, graphicx}
\usepackage{url, mhchem}

\newcommand{\reply}[1]{{#1}}

\begin{document}

\title{C/O Ratios and the formation of wide separation exoplanets}

\author[0000-0003-4179-6394]{Edwin A. Bergin}
\affil{Department of Astronomy,
University of Michigan, 1085 S. University Ave,  Ann Arbor, MI 48109, USA}

\author[0000-0002-0786-7307]{Richard A. Booth}
\affiliation{School of Physics and Astronomy, University of Leeds, Leeds, LS2 9JT, UK}

\author[0000-0000-0000-0000]{Maria Jose Colmenares}
\affiliation{Department of Astronomy,
University of Michigan, 1085 S. University Ave,  Ann Arbor, MI 48109, USA}

\author[0000-0003-1008-1142]{John D. Ilee}
\affiliation{School of Physics and Astronomy, University of Leeds, Leeds, LS2 9JT, UK}

\begin{abstract}
The gas and solid-state C/O ratios provide context to potentially link the atmospheric composition of planets to that of the natal disk.   We provide a synthesis of extant estimates of the gaseous C/O and C/H ratios in planet-forming disks obtained primarily through analysis of Atacama Large Millimeter Array observations.  These estimates are compared to atmospheric abundances of wide separation ($>$ 10~au) gas giants.  The resolved disk gas C/O ratios, from seven systems, generally exhibit C/O $\ge$ 1 with sub\reply{solar}, or depleted, carbon content.  In contrast, wide separation gas giants have atmospheric C/O ratios that cluster near \reply{or slightly above} the presumed stellar value with a range of elemental C/H.  %The lowest disk gas phase C/O ratios are also found in the disks believed to be in earlier stages of evolution.  
From the existing disk composition, we infer that the solid-state mm/cm-sized pebbles have a total C/O ratio (solid cores and ices) that is \reply{solar (stellar)} in content.   We explore simple models that reconstruct the exoplanet atmospheric composition from the disk, \reply{while accounting for silicate cloud formation in the planet atmosphere.}    If wide separate planets formed via the core-accretion mechanism, they must acquire their metals from pebble or planetesimal accretion.   Further, the dispersion in giant planet C/H content is best matched by a disk composition with modest and variable factors of carbon depletion.   An origin of the wide separation gas giants via gravitational instability cannot be ruled out as stellar C/O ratios should natively form in this scenario.  However, the variation in planet metallicity with a stellar C/O ratio potentially presents challenges to these models. 
\end{abstract}

%\keywords{tbd}

\section{Introduction} \label{sec:intro}

For the past decade there has been strong effort to explore the link between the exoplanetary atmospheric composition and the composition of its natal disk \citep[see][and references therein]{ObergBergin21}.    In part this has been motivated by the simple theory suggested by 
 \citet{omb11}.   This theory assumes interstellar abundances to focus on the main elemental carriers of C and O (CO, CO$_2$, and H$_2$O). The overall chemistry in the low ionization state disk midplane \citep{Umebayashi88, cleeves13a} dictates that sublimation alone is primarily responsible for changes in the ratio of carbon to oxygen in the ice versus gas as a function of stellar distance.   

\begin{figure*}

    \begin{center}
  \includegraphics[angle=0,width=1.0\textwidth]{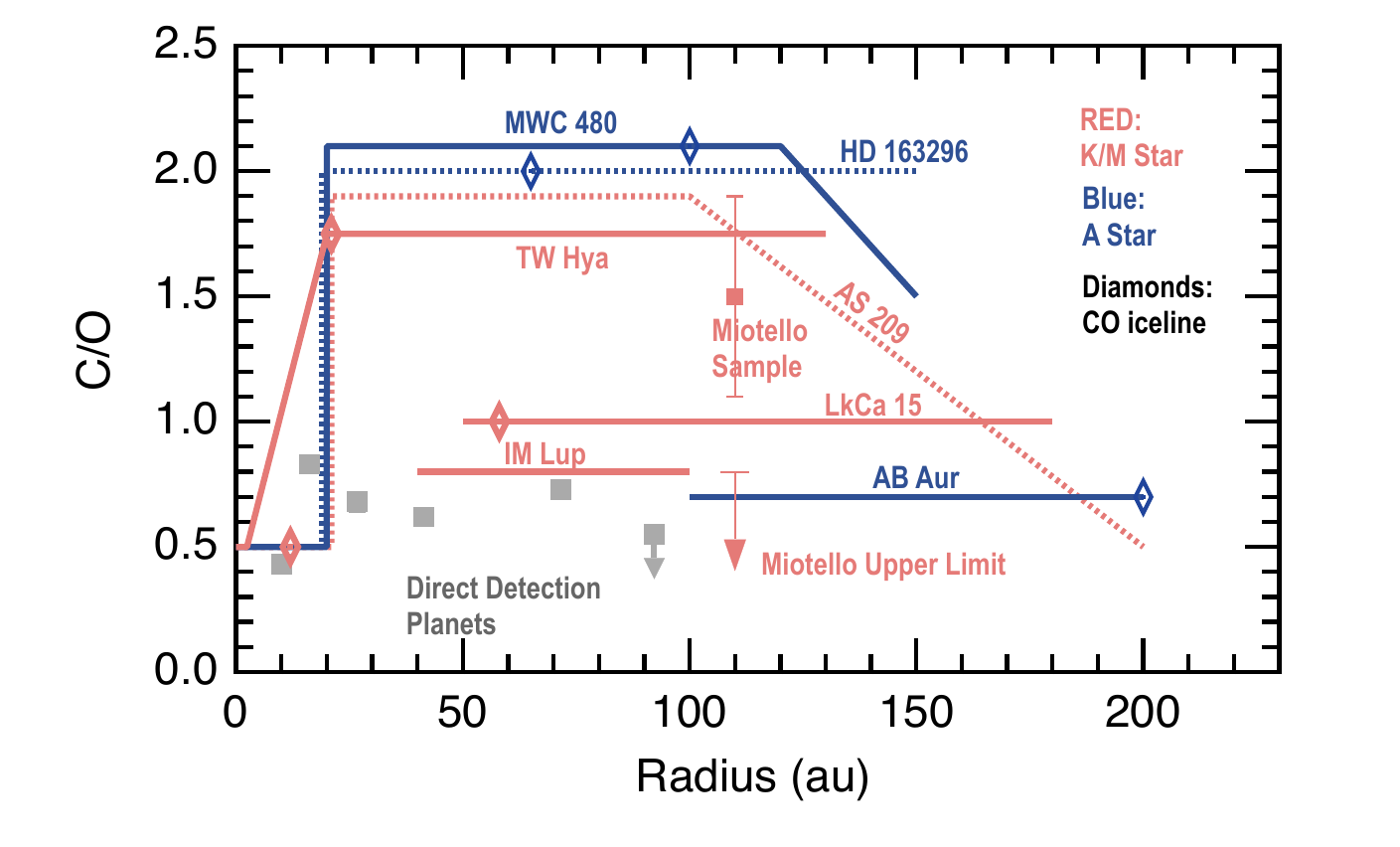}
  \vspace{-0.4cm}
   \caption{Radial distributions of derived C/O ratios in named disk systems from C$_2$H \citep{Bergin16, Kama16_co, Cleeves18, Bosman22, Sturm22} and CS/SO \citep{Riviere-Marichalar2022}.  Also shown is the range of C/O ratios estimated by \citet{Miotello19} in a survey of C$_2$H in Lupus disks.  This survey had some lower limits which imply smaller disks or lower C/O ratios.   In the figure A stars are shown in red and later spectral types in blue.
   The solid grey squares denote the estimated elemental C/O ratios towards direct detection planets $\beta$ Pic B \citep{Gravity20}, HR8799bcde \citep{Nasedkin2024}, HIP 65426 b \citep{Petrus21}.  \reply{The {\em estimated} error bars for some of the planet C/O ratios are smaller than the marker size. } The open diamonds superposed on a given C/O ratio line are the estimated location of the CO snowline for that object with references given in the Appendix.  For AB Aur the CO snowline is located at r $>$ 200~au and for IM Lup the CO snowline is near $\sim$15~au. }
   \label{fig:obsco}
   \end{center}
\end{figure*}

The relative balance of C/O between the gas and ice matters as within the core-accretion paradigm of giant planet formation, a many Earth-mass core forms from the icy solids.  
The H$_2$ dominated atmosphere {\em at birth} carries the composition of the gas at its formation distance.   The corollary is that ices, and the Earth-size core, are oxygen rich. This may be more complicated as planets migrate, can capture icy planetesimals, and might have core-atmosphere mixing along with gravitational settling \citep{Cridland16,Helled17, Guillot2022}. However, it encapsulates a central element and is now widely compared to exoplanet atmospheric composition retrievals \citep[][to list a few]{Barman15, Lavie17, Oreshenko17}.    As a corollary, an alternative route to formation would be direct gravitational collapse within an unstable disk \citep{Boss97, Durisen07}.   If this were to occur early, prior to significant grain growth within the disk, then a solar/interstellar composition would naively be predicted.

Observational constraints on the overall gas phase C/O ratio are obtained via analysis of data from the Atacama Large Millimeter Array (ALMA).
In gaseous emission lines of molecules ALMA is capable of resolving disk systems with $\sim$10-15~au resolution \citep{Oberg21, Law21} and numerous observational analyses have been undertaken to constrain the elemental C/O and C/H ratios in this gas \citep[][and references therein]{Miotello_ppvii}.
Many of these disk systems are suggested to be sites of incipient planet formation \citep[][and references therein]{Pinte_ppvii, Bae23_ppvii}.

Concurrently, new instruments (e.g. JWST) and ground-based high resolution spectroscopy have opened a new era in precision measurements of the C/O ratio in planets at wide distances from their stars encompassing similar spatial scales as probed by ALMA \citep{Gravity20, Wang20, Molliere20, Ruffio21, Petrus21}.
Thus, it is a fruitful time to revisit the current state of our understanding of  C/O measurements within young (1-10 Myr-old) gas-rich protoplanetary disk systems in comparison to the composition of exo-giant planet atmospheres.
In this paper we synthesize the observational state-of-the-art in linking disk and exoplanet atmospheric composition towards understanding the formation of wide separation planets in light of extant C/O ratio measurements.   We summarize existing {\em resolved} C/O ratio measurements to provide an observationally constrained plot of the C/O ratio as a function of distance moving beyond the simple theory of \citet{omb11}.  As part of this effort, we discuss the methodology applied toward retrieval of C/O ratios from gas phase emission lines.
Finally, we compare these results to extant measurements of C/O in the gas giant atmospheres and explore simple formation models to delineate what this comparison implies for the planet formation.
\reply{In this work we adopt a solar C/O ratio of 0.55 $\pm$ 0.06 as our reference frame and in Appendix A we outline how this value relates to the often uncertain stellar reference value for these systems. }

\section{ALMA C/O Measurements} \label{sec:c_o}

The methodology to determine the C/O ratio from ALMA observations of molecular line is outlined in Appendix B.
The majority of ALMA C/O ratio measurements are performed near and inside the CO snowline where the chemical expectation is that the carbon and oxygen in the gas are carried by CO, e.g. C/O = 1 \citep{omb11}.   The spatial coverage encompasses two zones within the disk, these are: (1) just inside the CO snowline where gas potentially traces the planet-forming midplane  and (2)  radii beyond the CO snowline, where gas CO is found only in warm (T $>$ 20--30~K) surface layers above the midplane where CO is presents as ice \citep{Aikawa02}.  For both regions the ratio can be indirectly measured using chemical systems that are both dependent on the elemental C/O ratio and have observable molecular transitions within the ALMA bandpasses.
The chemical systems are via the emission of C$_2$H and C$^{18}$O \citep{Bergin16, Kama16_co} and the ratio of CS/SO \citep{Semenov18, LeGal2021}.  

In Fig.~\ref{fig:obsco} we present the C/O ratio measurements in 7 disk systems surrounding A stars (3) and K/M stars (4).  
In Appendix C we provide the source by source discussion of this measurement from molecular emission images.  \reply{For simplicity we adopt solar abundances as our reference frame with a C/O ratio of $\sim$0.55; this is discussed in Appendix A.}
Fig.~\ref{fig:obsco} also shows the estimated CO snowline for resolved systems.  The other relevant snowline on these spatial scales is CO$_2$. Assuming a CO$_2$ sublimation temperature of $\sim$55~K \citep{Minissale22}, we estimate (rough) CO$_2$ snowline locations of $\sim10-15$~au for the Ae/Be systems (HD 163296 and MWC 480) and inside 3~au for the T Tauri disks (TW Hya, IM Lup, and AS 209).  The CO$_2$ snowline lies inside the dust cavities of LkCa15 and Ab Aur.  
Based on this information the majority of measurements are nominally tracing material where the expectation value is C/O = 1 and ALMA measurements retrieve near this value in 3 instances (LkCa 15, IM Lup, and AB Aur) but in other disks the gas phase C/O is estimated to be $> 1$ beyond 20~au.  Inside 20 au a lower ratio is inferred.

In Appendix D we also summarize estimates of the  gas phase CO abundance in these disks which appears to be reduced compared to interstellar by $\sim$10 or more \citep[see source by source discussion in the Appendix and also][]{Bergin17, Miotello_ppvii}. Since CO comprises nearly 50\% of the available carbon \citep{Bergin15, Mishra15} this is well below the expected value.  Beyond the CO$_2$ snowline the abundance of CO is believed to trace the C/H content of planet-forming gas \citep{omb11}; thus, to date, most planet-forming disks have C/O $\gtrsim 1$ and  C/H $<$ C/H(\reply{solar}) \citep{Bosman21_mapsco}. 

%CO snow: 
%IM Lup 15, AS 209 12, HD 163 65, MWC 100 (from Zhang et al. Maps, ~20 K),  AB Aur (42 K @ 98 au from Riviere-Marichalar so 300 au) 300 au  LkCa15  58 au (Qi et al. 2019; Sturm et al. 2023)  TW Hya  21 au (Schwarz et al. 16; Zhang et al. 2017)
%CO2 snow: scale it via r^1/4 power; minisale gives T_sub(CO2) near 60-70K so let’s say 65 K. IM Lup 2 au; AS 209 1.2 au; HD 163 6.5 au; MWC 10 au; AB Aur 30 au; LkCa 15 5.8 au;  TW Hya 5 au (Bosman et al. 2019).

\section{Planet Formation and C/O} \label{sec:pf}

\subsection{General Implications}

\begin{figure}[t]
    \begin{center}
\includegraphics[angle=0,width=0.5\textwidth]{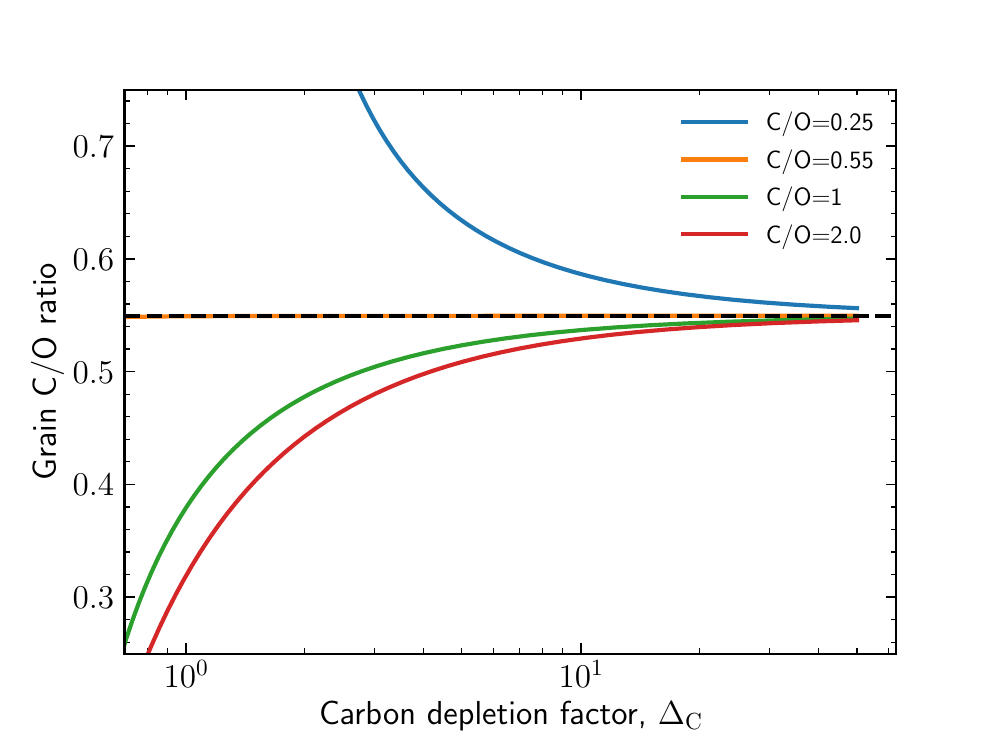}
  \vspace{-0.4cm}
   \caption{Estimated C/O ratio within the grain icy mantle and solid core as a function of the gas phase C/O ratio and the overall carbon depletion factor.  Here we assume a solar composition for the stellar content. The dashed line represents the solar/stellar C/O ratio.}
   \label{fig:solidco}
   \end{center}
\end{figure}

Fig.~\ref{fig:obsco} also provides estimates of the atmospheric C/O ratio of wide separation gas giants illustrating a general mismatch between the C/O ratio measured in direct detection gas giants and the gas phase C/O in natal disk systems \reply{(discussed in Appendix~\ref{sec:exo_CO}).}  
Placing the focus on planet and disk material beyond 20 au the majority of disk systems subject to detailed chemical analyses have gas phase C/O $\ge$ 1 while planetary material appears closer to \reply{solar} values.   
Another clear statement can be made from Fig.~\ref{fig:obsco}: if the \reply{disk} gas phase C/O is greater than \reply{solar}, and C/H is depleted by a factor of $\sim$10, then the ice coatings of the pebbles likely have solar composition.  This is illustrated in Fig.~\ref{fig:solidco} where we plot an estimate of the C/O ratio of the pebbles as a function of the overall depletion factor of carbon (traced by CO) and the gas-phase C/O ratio. Specifically, we define the depletion factor $\Delta_\mathrm{C}$ relative to the volatile carbon content in the ISM (assumed to be half the solar carbon abundance, \citealt{Mishra15}), such that $\Delta_\mathrm{C} = 2$ implies that $1/4$ of the carbon is in the gas phase.  If C/O $\ge$ 1 and $\Delta_\mathrm{C}\sim$10, then the icy grain mantle C/O ratio is close to solar in composition.   That is the grains contain the majority of the solids (silicates, carbonaceous materials) and the volatiles (e.g. CO, H$_2$O, and CO$_2$).

There are two rough groupings of C/O within the disk systems with C/O = 1.5--2.0 (AS 209, HD 163296, TW Hya, and MWC 480) and those with C/O = 0.7--1.0 (LkCa 15, Ab Aur, IM Lup).    This difference {\em could} be due to evolutionary differences as the latter sources are relatively young (1-5~Myr) while the former are generally older ($>$ 5~Myr).  However, this is not completely the case.  AS 209 has been subject to two semi-independent chemical analyses \citep[i.e., two different codes using the same baseline physical structure;][]{Bosman21_mapsco, Alarcon21} has an elevated C/O ratio and is estimated to be a young (1-3~Myr old) system.   Abundance evolution is inferred to be present for CO, the primary gaseous C/H carrier, over million year timescales \citep{Bergner19,Zhang19}, and the elevated C/O ratios may develop over similar timescales.
Thus, chemical evolution with planet formation within a few Myr is one possibility.   

\begin{figure*}[t]
    \begin{center}
  \includegraphics[angle=0,width=1.0\textwidth]{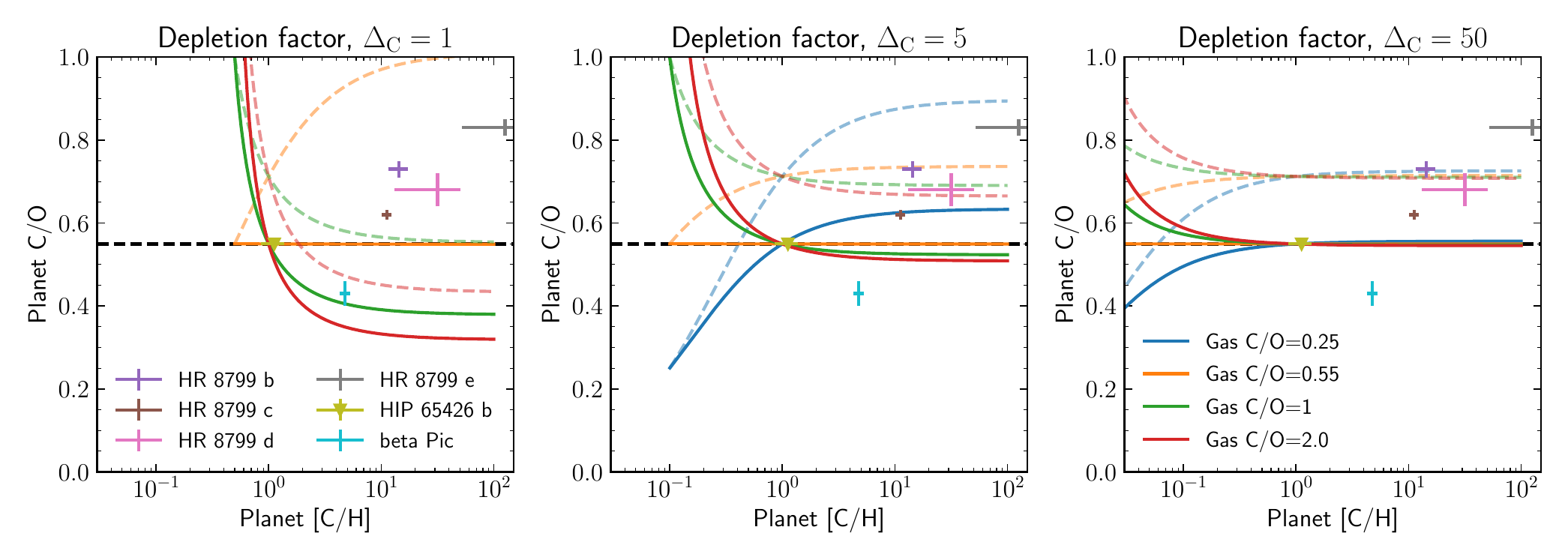}
  \vspace{-0.4cm}
   \caption{A comparison between the observed composition of planets (crosses) and the compositions obtained by combining different amounts of solids and gases (lines). \reply{The solid lines show the potential planet compositions for a given disk C/O ratio; in different panels, we have varied the amount of carbon depletion as in Fig.~\ref{fig:solidco}. The solid lines assume all material contributes to the planet's composition, while the dashed lines assume that silicates condense out.} For HIP~65426~b, the C/O ratio shown is the upper limit \citep{Petrus21}.}
   \label{fig:planets_soalar}
   \end{center}
\end{figure*}

\subsection{The Focus on A Star Disks}

An additional aspect is that the direct detection host stars are primarily A stars.  In this case the comparison sample is HD 163296, MWC 480 and AB Aur.  Again, age may play a factor as AB Aur is the youngest system \reply{and displays C/O $\sim$ 0.7--1 (see Appendix C.3).}  In fact, Ab Aur is suggested to be accreting material from its natal cloud \citep{Tang12}  and also hosts a potential protoplanet \citep{Currie22}.  The accretion of this fresh material may alter the C/O ratio in the disk gas.   In general, we expect that material accreted from the surrounding cloud will have a composition consistent with interstellar gas and ices. These have CO in the gaseous state with most oxygen confined to the refractory cores of grains and in CO, CO$_2$, and water ice \citep{oberg11_c2d, McClure23}.  To lower the C/O ratio below unity some of the oxygen trapped in CO$_2$ and H$_2$O ice needs to be released to the gas.   There is some evidence of chemical changes associated with accretion in protostars which have strong SO  and SO$_2$ emission sometimes spatially associated with a young protostellar disk \citep{Sakai14, ArturdelaVillarmois22,Flores23, Kido23}.  The presence of these oxygen-rich sulfur-bearing species is suggestive that some oxygen is returned to the gas, a facet that is consistent with models \citep{Miura17, vanGelder23}.  

However, we have no information regarding the C/O ratio for Ab Aur inside 100 au.  Thus, it is not clear that age is an issue.  The systems where we do have information that is spatially coincident with the direct detection exoplanet population is MWC 480 and HD 163296.  In both systems beyond 20~au the gaseous C/O ratio is elevated near 2 and well above the values inferred in the exoplanet atmospheres.  Inside 20~au the situation is less clear.   HD 163296 provides the best information via the rotational emissions of CH$_3$CN and HC$_3$N which peak near 40~au \citep{Ilee21} and decay towards smaller radii.  This is interpreted by \citet{Calahan22maps} as a reset in the C/O ratio, but this is uncertain.  What is clearer is the fact that gaseous C/O ratio beyond 20 au in most disk systems appears to be elevated above that measured in the direct detection planets.

\begin{figure*}[t]
    \begin{center}
  \includegraphics[angle=0,width=1.0\textwidth]{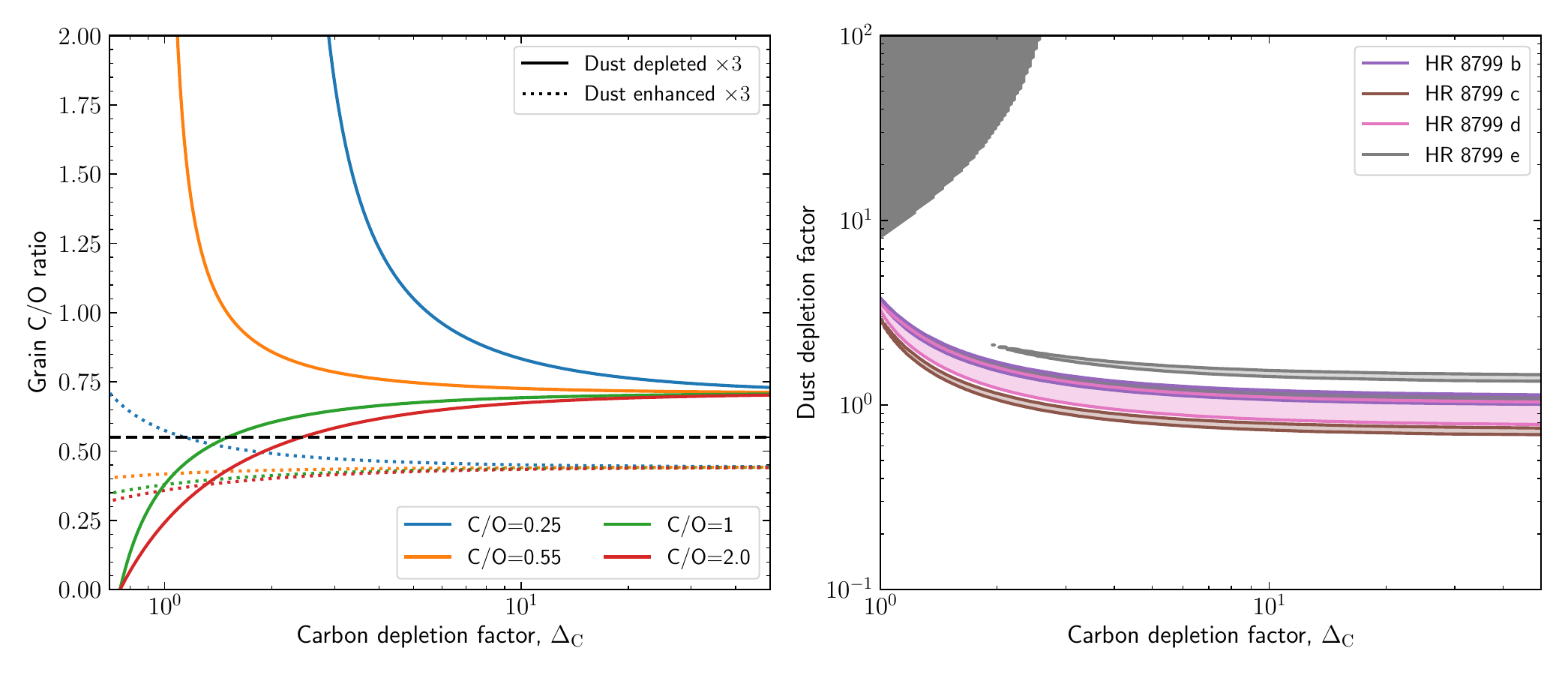}%{Grain_abundance_depleted_CO_0.5.pdf}
  \vspace{-0.4cm}
   \caption{Left: Estimated C/O of the dust, as in Fig~\ref{fig:solidco} but considering a depletion (solid lines) or enhancement (dotted lines) of icy material before chemical evolution alters the disk composition.  \reply{The black line represents the solar ratio.} Right: Constraints on the level of carbon depletion and dust depletion provided by the planets in the HR~8799 system, assuming the gas has a C/O ratio of 2. \reply{The filled regions in the contours show the regions where the disks can produce planets that match the observed composition to within $1\sigma$, under the assumption that silicates condense into clouds and do not contribute to the observed composition.} Note that a dust depletion smaller than one means that the dust-to-gas ratio has been increased. The other planets in Table~\ref{table:exo} provide similar, but weaker constraints. }
   \label{fig:solidco_dep} \label{fig:planets_dep}
   \end{center}
\end{figure*}

\section{Discussion}

Current estimates clearly suggest a mismatch between the gas-phase C/O ratios measured in disk systems and that in exoplanet atmospheres.   Below we focus on simple models of planet formation that encompass the disk gas phase C/O and C/H constraints (with the implication that constraining both of these quantities in the gas also implies an O/H ratio) and explore how these might be consistent with existing exoplanet atmospheric constraints.

\subsection{Modes of Planet Formation} 
\label{sec:modes}

We consider that the composition of the wide orbit giants is likely controlled by the composition and relative amount of gas and dust accreted. Using the gas abundances derived for the outer regions of protoplanetary disks and the compositions of wide-orbit giants (see Table~\ref{table:exo}) that likely formed in these regions, we can begin to understand how these planets may have formed. 

\reply{Our model methodology is given in Appendix F.}
We first consider the case where the total abundances of the gas and solids in the disk add up to the stellar abundances, for which we use the proto-solar abundances \citep{asplund09} as a proxy. Partitioning the abundances based on the C/O ratio of the gas and the amount of carbon depletion, we then compute possible compositions for the planets by varying the amount of solid material accreted, with the results for typical C/O ratios and carbon depletions shown in Fig.~\ref{fig:planets_soalar}. Increasing the amount of solids in the planet increases the planet's carbon abundance [C/H] and drives the C/O ratio towards that of the solids. In these solutions, disks with significant carbon depletion are generally preferred. \reply{In drawing this conclusion, we have accounted for the condensation of silicate clouds in the HR 8799, which raises the C/O ratio of the atmosphere because oxygen is locked up in silicates. See Appendix~\ref{sec:exo_CO} and \citet{Nasedkin2024}, for a discussion.} %This is because the planets have a wide range of C/H ratios, but C/O ratios that are close to solar.  This requires some variability in the C/H disk content. 
A corollary is the planets must have acquired most of their metals through the accretion of solids. The one exception is $\beta$~Pic~b, for which the high C/H and low C/O ratio prefer a low carbon depletion.

Increasing the amount of solids in the planet increases the planet's carbon abundance [C/H] and drives the C/O ratio towards that of the solids. In these solutions, disks with significant carbon depletion are generally preferred.  This is because the planets have a wide range of C/H ratios, but C/O ratios that are close to solar.  This requires some variability in the C/H disk content. A corollary is the planets must have acquired most of their metals through the accretion of solids. The one exception is $\beta$~Pic~b, for which the high C/H and low C/O ratio prefer a low carbon depletion.

Alternatively, the HR~8799 and HIP~65426~b planets' composition could be explained by any level of carbon depletion if the C/O ratio of the gas is close to the solar value. In this case, both the dust and gas have similar compositions, and the only planet formation outcome possible is a C/O the solar value. Only AB Aur has a C/O ratio close to solar, however, and this is the youngest disk. This suggests an alternative explanation for the composition of the planets, which is that they formed early, before chemical evolution was significantly underway.

\begin{figure}
    \begin{center}
  \includegraphics[angle=0,width=0.45\textwidth]{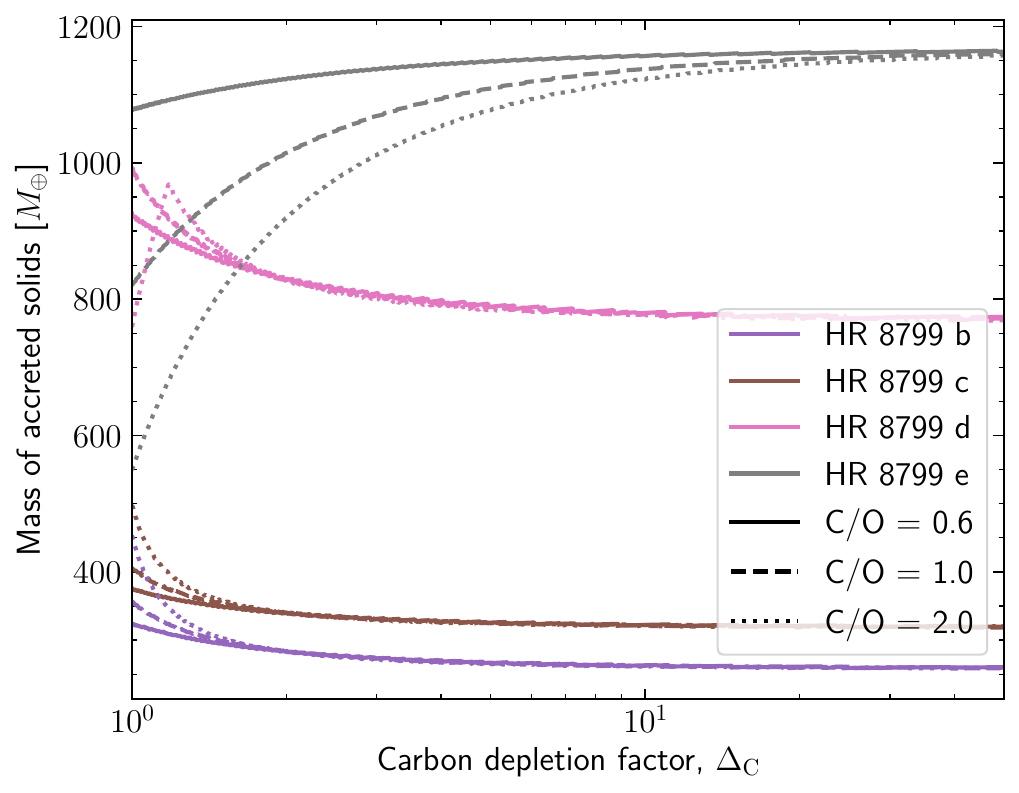}
  \vspace{-0.4cm}
   \caption{Estimated mass of accreted solids from simple model as a function of the carbon depletion factor ($\Delta_C$) and the dust depletion factor. \reply{The dust depletion factor was chosen to be the best fit value based on the C/H of the planets, and it corresponds to $\sim$0.5--5, depending on the assumed C/O ratio. When the dust depletion factor is less than one it reflects a local dust enhancement.}}
   %When the latter value is less than one it reflects a local dust enhancement.}
   \label{fig:planets_Macc}
   \end{center}
\end{figure}

\subsection{Gas and Dust Separation}

Since the composition of ice in these disks is not known, we cannot be certain that the total abundance of gas and solids is solar. A non-solar total abundance requires that the dust and gas evolve separately, but processes such as radial drift or dust trapping in sub-structures can affect the disk composition by removing or enhancing the amount of ice \citep{Pinilla17, Kalyaan23}.   Based on ALMA observations, it is clear that in many disks the outer gaseous radii are much larger than the mm-sized dust \citep{Ansdell18, Sanchis21} which demonstrates that some gas exists in regions where dust is depleted.
This would not alter our conclusions above if the dust evolution happens after the chemical evolution. However, there is good evidence that the dust could evolve more quickly. For example, sub-structures are seen at all ages \citep{Huang18}, and theory predicts that dust will evolve rapidly in their absence.  

To estimate how differences in the gas and dust evolution affect the composition of the solids we need to know how much material is lost together with its composition. We assume that any dust evolution happens while the disk has its initial composition; this choice maximizes the impact of dust evolution if the gas composition evolves monotonically \reply{everywhere in the disk}. The initial composition is assumed to be inherited from the interstellar medium, in which approximately 50\% of the carbon is in the form of CO \citep{Mishra15}. In the cold outer regions probed by the ALMA observations, all other major carbon and oxygen carriers will be in the form of ice. From this, we can estimate the carbon and oxygen abundance of the ice by first computing a new total abundance from the gas-phase CO abundance plus the dust abundances divided by a {\em dust} depletion factor (which could reflect a true depletion or an enhancement in a trap). The final dust abundance is then computed by assuming this new total composition is partitioned between the gas and dust according to the observed C/O ratio and carbon depletion factor. We show how this affects the composition of the grains in Fig.~\ref{fig:solidco_dep}. When dust enhancement occurs (e.g. via dust trapping), the final composition of the grains is closer to their ISM abundance ($\mathrm{C/O}\approx 0.38$) as the gas is less important. Dust depletion has the opposite effect, with the dust's initial contribution being smaller, increasing the dust's C/O ratio towards unity (since the gas contained only CO). 

Ultimately, it is likely that the C/O ratio of \reply{protoplanetary disk} dust is not too different from solar when most of the planets formed. \reply{Our models predict only a small change in the solid C/O ratio of the disk for reasonable changes in the amount of dust ($\lesssim 0.1$ for $0.3 \lesssim \Delta_{\rm d} \lesssim 3$, Fig.~\ref{fig:planets_dep}, left panel), and the inferred compositions of the planets in HR 8799 favor little-to-no change in the amount of dust (Fig.~\ref{fig:planets_dep}, right panel).} While we have only included HR~8799 \reply{planets, HIP~65426~b provides} consistent, albeit weaker, constraints. This confirms our previous statement that the planet's compositions prefer a modest level of carbon depletion in the disk unless the planets formed early, when the disk had a low C/O \reply{prior to significant dust evolution.} $\beta$~Pic~b is again an exception in this regard, with the planet's low C/O and high C/H ratios either favoring no depletion and a gas phase $\mathrm{C/O}\approx1$ (Fig.~\ref{fig:planets_soalar}) or a dust enhancement.

\subsection{Metal Enrichment of Planet Envelope}

Giant planet formation models based on planetesimal accretion, pebble accretion, and gravitational instability differ in their predictions for the amount of solids accreted.
In Fig.~\ref{fig:planets_Macc}, we show the amount of solids \reply{the HR 8799 planets must have accreted to match the metallicity and C/O ratios inferred by \citet{Nasedkin2024}, assuming the best-fit dust depletion factor for each $\Delta_{\rm C}$. These high metallicities inferred for the HR~8799  planets by \citet{Nasedkin2024} imply the accretion of several 100~$M_{\oplus}$  to 1000~$M_{\oplus}$ of solids to match the observed composition. The total solids in all of the HR~8799 planets exceed 2000~$M_{\oplus}$ and is comparable to the amount of dust in a solar mass of solar-composition gas.} 

\reply{It is unlikely that any standard planet formation scenario can explain these abundances. Pebble accretion is expected to stop once a planet reaches the pebble isolation mass and the pebble isolation mass is unlikely to exceed $\sim 100\,M_{\oplus}$ \citep{Bitsch2018}, ruling out pebble accretion as a mechanism to explain these metallicities. Similarly, there is unlikely sufficient mass in planetesimals to explain the metallicities. While gravitational instability can also produce super-solar metallicities if dust concentrates in the spirals before collapse \citep{2010ApJ...724..618B}, metallicities above a few times solar have not yet been seen in simulations. As a result, no known formation pathway naturally explains the high metallicities inferred for these massive planets.} 

\reply{Estimates of the solid abundance based on previous studies \citep[e.g.][]{Molliere20,Ruffio21,Wang20} produce abundances that are a factor $\sim 10$ lower \citep{Molliere20} due to the lower metallicity ($\lesssim 3$ times solar) inferred by these studies. These lower abundances are less problematic, and each formation scenario could likely achieve them.}

\reply{Putting aside the question of how the very high metallicities can be achieved, we address what the planets' compositions tell us about how the planets formed. Fig.~\ref{fig:planets_dep} tells us that the HR 8799 planets likely formed in a disc with a modest level of gas-phase carbon depletion and solids that had a composition close to solar if silicates are condensing in the planet's atmospheres.} 

\reply{Planetesimal accretion models are consistent with the C/O ratios in the HR~8799 planets when considering the carbon depletion seen in disc observations.} The main challenge for planetesimal accretion would be explaining the overall metallicity and mass-metallicity trends in the HR~8799 system. Planetesimal accretion models predict anti-correlated mass-metallicity relations \citep[e.g.][]{Thorngren2016,Mordasini16}, but the HR~8799 planets do not follow this.

\reply{The main challenges for forming the HR 8799 planets by pebble accretion are}  1) explaining how the pebbles ended up enriching the planet's atmosphere since they would have been accreted while the planet was less than the pebble isolation mass and thus before the planet accreted most of the gas. These pebbles would initially form a dilute core that would have to be mixed into the atmosphere to contribute to the current day planet abundances \citep{Ormel2021}. 2) Demonstrating that giant, metal-rich planets can form without the drifting pebbles enriching the disk gas, since large enrichments produce planets with $\mathrm{C/O}\sim 1$  and disks with [C/H]$\gtrsim 1$ \citep{Booth2017, Booth19, Danti2023}.

\reply{It is also difficult to reconcile the HR 8799 planets' abundances with formation via gravitational instability since the abundances require some level of carbon depletion or dust depletion (Fig.~\ref{fig:planets_dep}) and this is not expected to occur in the short period of time when gravitational instability is viable. However, gravitational instability takes place in an extremely dynamic disc, which can affect the disc's composition \citep{2017MNRAS.472..189I}; the impact of these effects on the planets' composition are currently unknown.} Condensation and settling of grains within a fragment would affect the composition too, as would the subsequent removal of the gaseous envelope (so called tidal stripping or downsizing) would also result in objects with higher metallicity \citep[][]{2010ApJ...724..618B, 2010MNRAS.408L..36N} and different elemental ratios \citep[e.g.][]{2017MNRAS.472..189I}.

\reply{Here, we have focused on the composition of the HR~8799 planets because they provide the most challenges. Of the other planets considered, HIP 65426~b has been inferred to have a composition close to solar, which is compatible with essentially all formation scenarios. $\beta$~Pic~b would also likely be compatible with any formation channel if it formed early, when $\Delta_{\rm C} \approx 1$ and ${\rm C/O} = 1$ are expected in the outer parts of the disc.}

\section{Summary} \label{sec:summary}

We present a synthesis analysis of the C/O ratio in the outer regions of gas-rich protoplanetary disks in comparison to that found in wide separation exoplanet atmospheres. This comparison leads towards a number of conclusions.

\begin{enumerate}
    \item  The disk resolved gaseous C/O ratio from seven disk systems has C/O $\ge$ 1 with subsolar C/H content.   These measurements correspond to disk locations where the baseline chemical expectation is C/O = 1.  
    \item Within this sample the youngest sources appear to have the lowest C/O ratios which tentatively (small sample of systems) hints at chemical evolution in the overall C/O ratio, which may be commensurate with similar evolution in C/H \citep{Bergner19, Zhang19}.  
       
    \item Based on the gas phase C/O and C/H values inferred for the gas we conclude that the total C/O ratio of pebbles in planet-forming disks, including the solid cores and the ice mantles, have a C/O ratio that is \reply{solar} in content.   The results from the ALMA large program ``The Disk Exoplanet C/Onnection'' (DECO), will provide the needed statistics to confirm these conclusions.
        
    \item Disk C/O ratios are uniformly above that measured in wide separation exoplanet atmospheres where the ratio is closer to solar or stellar (where known). Using a simple analysis based on the constraints of the gaseous and solid state composition we show that the exoplanet composition can be matched provided that pebbles provide the bulk of planet metals.   As an additional constraint the dispersion in giant planet atmospheric carbon content (e.g. C/H) requires a modest level of carbon depletion in disk gas.
    
    \item Based on this comparison we cannot conclude on the primary method of planet formation whether gravitational instability or core accretion.   However, we do note that the youngest disk sources have the lowest C/O ratios.  This  may favor a solution where planets form early before significant chemical evolution ensues.  However, if the range in the C/H content in exoplanet atmospheres is correct then the dual constraints of requiring \reply{solar}/stellar C/O and variable C/H challenges some solutions.
\end{enumerate}

%\begin{acknowledgments}
We are grateful to a through and constructive review from an anonymous referee which improved this manuscript.
E.A.B. acknowledges support from NSF grant No. 1907653 and NASA NASA's Emerging Worlds Program, grant 80NSSC20K0333, and Exoplanets Research Program, grant 80NSSC20K0259. R.A.B. thanks the Royal Society for their support via a University Research Fellowship. J.D.I. acknowledges support from an STFC Ernest Rutherford Fellowship (ST/W004119/1) and a University Academic Fellowship from the University of Leeds. 
%\end{acknowledgments}

\vspace{5mm}
\facilities{ALMA, NOEMA}


\software{astropy \citep{2013A&A...558A..33A, 2018AJ....156..123A},
          }



\appendix

\section{Interstellar Reference C/O vs Stellar C/O}

\reply{All of the young disks in our discussion are within $\sim$150~pc and have formed from local interstellar medium gas.  The C/O ratio of this gas has been estimated through the determination of photospheric abundances of early type B-stars in OB associations \citep{Nieva12} and this reference value is C/O = 0.37$\pm$0.06.  This is in comparison to the solar value of 0.51$\pm$0.06 using the most recent solar oxygen \citep{Bergemann2021} and carbon \citep{Asplund21} abundance estimates.  The solar value represents the local ISM circa 4.6 Billion years ago and the differences potentially represent galactic chemical evolution in that time span.  However, the error estimates between the interstellar value and the solar value overlap and for our purposes we adopt the solar value as a reference value.  Further, in this work we use the solar composition reported in \citet{asplund09} with C/O = 0.55 $\pm$ 0.06 as this value is widely used in the literature and its use provides a common reference point for comparison to earlier works.  

The direct detection planets stellar C/O ratios are not always known. However, \citet{Wang20} estimate a stellar C/O ratio of 0.54$^{+0.12}_{-0.09}$ for HR8799.  More generally, \citet{Biazzo22} surveyed the stellar abundances of transiting exoplanet host stars finding that all host stars have C/O $<$ 0.8 with peak values in two metallicity bins, $Z_\star/Z_\odot$ $\le$ 1.3 and $>$ 1.3, of $\sim$0.45 and 0.5, respectively.  In sum, while there is uncertainty, the expectation is that the stellar value is close to 0.5 or slightly below.  Certainly the variation in the estimated C/O ratio in HR8799 \citep{Nasedkin2024} seen in Fig.~\ref{fig:obsco}, if supported by future work, is suggestive of planetary deviations from stellar.
}

\section{Outline of Methodology to Derive C/O from ALMA Observations}

\subsection{\ce{C2H} and C$^{18}$O}

C$_2$H emission is noted as unusually strong in disk systems with emission levels rising to be commensurate with $^{13}$CO in some instances \citep{Kastner14}.  
The C/O ratio is estimated using this tracer via forward-modeling of individual systems through detailed (thermo-)chemical models.
The first step  involves fitting the overall dust spectral energy distribution from near-IR to mm wavelengths alongside the resolved flux distribution of the dust sub-millimeter emission within the framework of known stellar parameters (accretion rate, stellar mass/radii, and luminosity) with parametric models of the distribution of  the dust density both radially and vertically \citep{Andrews20araa}.  Radiation transfer within the mass distribution, with assumed dust properties \citep{Pollack87, Birnstiel_dsharp}, sets the temperature distribution of the dust disk.  Based on observations, the majority of the dust mass resides in the midplane held by larger mm-sized grains with smaller grains coupled to the gas following the flaring of the gaseous disk set by hydrostatic equilibrium \citep{Dutrey17, Villenave20}.   Thermochemical models, which simultaneously simulate the chemistry and molecular line cooling, are used to solve for the thermal and chemical properties of surface layers where the dust and gas temperatures diverge \citep{woitke09, gorti11, Bruderer12, Du14}.  

Models generally first match the distribution of C$^{18}$O.
In this framework the gas mass and the CO abundance are degenerate \citep[e.g.,][]{Calahan21twhya}. The overall abundance of CO is an important issue and is effectively set via assumptions about the disk gas mass.  In these models, the CO abundance  effectively sets C/H (and O/H) in the gas beyond the \ce{CO2} iceline.  In this paper we will report CO abundances as measured in the given paper and refer the reader to the summaries in \citet{Bergin17} and \citet{Miotello_ppvii}.   With the CO abundance set, the overall elemental C/O ratio is varied by adding excess carbon (in the form of C I or \ce{CH4}) and matching the level of \ce{C2H} emission and/or estimated column density.  This gives leverage for the C/O ratio from the assumed value of 1 and above.  Since \ce{C2H} is a tracer of UV illuminated gas \citep{Nagy15} there is an additional dependence on the amount of small grains present in surface layers \citep{Bosman21_mapsco}.  There are two variations to this method.  \citet{Cleeves18} traced C/O ratios below unity by adding additional water ice into the model which can be photodesorbed to provide gas phase oxygen to destroy \ce{C2H}.  Finally, \citet{Calahan22maps} extended tracers of C/O to include the overall complex carbon chemistry through the emission of \ce{CH3CN} and \ce{HC3N}; HCN is also used \citep{Cleeves18, Riviere-Marichalar2020}.   For additional information the reader is referred to the discussion in \citet{Fedele20}.

\subsection{CS and SO}

The CS/SO ratio in existing measurements also currently probes gas beyond the \ce{CO2} snowline.   This chemical system has long been posited as a sensitive probe of the C/O ratio in dense interstellar medium gas \citep{Bergin97_csso, Nilsson00} as oxygen rich gas readily forms SO, while carbon-rich material favors the formation of CS over SO.   \citet{Semenov18} and \citet{LeGal2021} demonstrate that this ratio maintains its effectiveness as  a probe of C/O in disk systems.   The methodology for the derivation of C/O is similar to that of \ce{C2H} in terms of model development with one key caveat: in comparison to \ce{C2H} the ratio of CS to SO is mass-independent requiring only detection or limits on the emission of CS and SO. 
We note that in some instances there are indications of non-axisymmetric structure in the emission of SO, hinting at localized variations in the C/O ratio \citep[e.g.,][]{abooth23, Keyte23}. In this paper we discuss the majority of systems where, at present, emission appears to be symmetric and tracing the generic state of planet-forming material.

\subsection{Comparison of Methods}
\label{sec:diskcomp}

 Of the disk C/O ratio estimates shown in Fig.~\ref{fig:obsco}, AB Aur is the only one to have its C/O determined via CS/SO. All other C/O estimates were determined using C$_2$H and CO.  Thus, it is possible that there could be a systematic effect between the two different methodologies.    In this sample the only example where both have been used is towards MWC 480 where the spatial distribution of gas-phase CS and C$_2$H is constrained \citep{Law21, Guzman21, LeGal2021}, alongside an upper limit to the emission distribution of SO \citep{LeGal2021}.   In this regard the CS/SO limit is consistent with C/O $>$ 0.9 in the same gas where C/O is estimated to be $\sim$2 from C$_2$H. 
 Thus, there is some baseline consistency, but clearly more work in this space is needed. 

\section{Description of Source Specific C/O Measurements}

\subsection{HD 163296}

\begin{figure}
    \centering
    \includegraphics[width=0.7\textwidth]{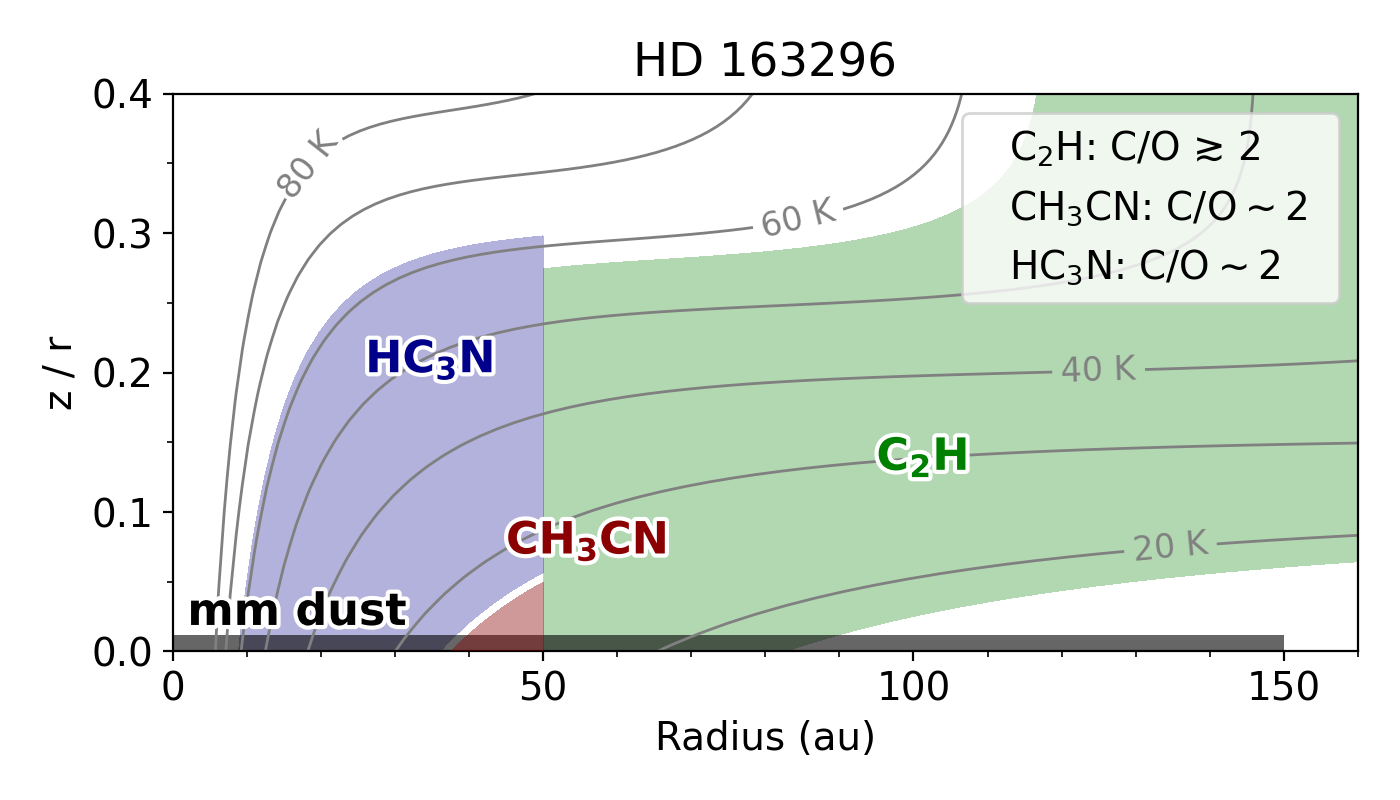}
    \caption{\reply{Origin of the line emission in the HD~163296 disk as determined in \citet{Guzman21} and \citet{Ilee21}. Also shown are the approximate C/O ratios derived from the analysis of \citet[][from C$_2$H]{Bosman21} and also \citet[][from CH$_3$CN and HC$_{3}$N]{Calahan22maps}.}}
    \label{fig:hd163_line_origin}
\end{figure}

HD 163296 is an A star (M $=$ 1.9$\pm$0.1~M$_\odot$) with a luminosity of 17~L$_\odot$ located at a distance of 101~pc \citep{Fairlamb15, Gaia2018}.  The age of this system is estimated to be of order 5-7~Myr \citep{Montesinos09, Fairlamb15}.   The C/O ratio in this system is measured by \citet{Bosman21_mapsco} by matching C$_2$H and C$^{18}$O emission.    The ratio is estimated to be $\ge$ 2 beyond 50~au.  Inside 50~au there is a decrease in the \ce{C2H} column.  This rise does not appear to occur at the CO ice line, provided that its sublimation temperature is 22~K \citep{Harsono15, Bosman21_mapsco}.   \citet{Calahan22maps} use \ce{CH3CN} and \ce{HC3N} emission to estimate C/O in the inner 50~au for HD~163296.  They show that the C/O ratio is comparable to C$_2$H from 20-50~au but must decline in the inner 20~au. We follow their suggestion of C/O = solar in this gas. \reply{Figure \ref{fig:hd163_line_origin} gives an overview of the radii and relative heights in the disk that are probed by these observations.  This figure illustrates that, for this system at least, the elevated C/O ratios extend closer to the star and trace both surface layers (HC$_3$N and C$_2$H) and material near the midplane as CH$_3$CN emission must arise from cold gas deep inside the disk \citep{Ilee21, Guzman21, Calahan22maps}.}  The CO abundance (C/H) beyond 100~au is of order 10$^{-5}$ \citep{Zhang21_mapsco} and potentially increases to super-\reply{solar} in the inner tens of au \citep{Zhang20_hd163, Zhang21_mapsco}.   The CO snowline is estimated to lie at 100~au by \citet{Zhang21_mapsco} based on CO isotopologue emission.

\subsection{MWC 480}

MWC~480 or HD~31648 is an A5-A6 star (M = 1.85$^{+0.04}_{-0.01}$~M$_\odot$) with a luminosity of 16.6~L$_\odot$ located at a distance of 162 pc \citep{Gaia2018, Guzman-Diaz21}.  The age of this system is estimated to be 6-7~Myr \citep{Montesinos09, Guzman-Diaz21}.    The C/O ratio in this system is measured by \citet{Bosman21_mapsco} by matching C$_2$H and C$^{18}$O emission \reply{to find} C/O $\ge$ 2 beyond 40~au.  The \ce{C2H} column exhibits a strong decline beyond 120~au which is interpreted as a change in the C/O ratio, and we place a smooth gradient in the plot to trace this decline. The C/O gradient beyond 120~au and this interpretation are highly uncertain.   In the inner 40 au the \ce{C2H} column exhibits a decline.   As in HD~163296 this decline lies inside the estimated location of the CO snowline.  We place the reset of the C/O ratio to \reply{solar} near 20~au based on the strong \ce{CH3CN} emission inside 40~au \citep{Ilee21}.  The late stage chemistry with an elevated C/O ratio invoked by \citet{Calahan22maps} for HD~163296 is likely relevant for MWC~480.   Thus we assume that a shift to lower C/O occurs closer to the star inside 20~au.
The CO abundance (C/H) beyond 80~au is of order $\sim$5 $\times$ 10$^{-6}$  and potentially increases to solar or super-\reply{solar} in the inner tens of au \citep{Zhang21_mapsco}.   The CO snowline is estimated to be found near 65~au by \citet{Zhang21_mapsco} based on analysis of CO isotopologue emission.

\subsection{AB Aur}

AB Aur is an A1-A2 star (M = 2.36$^{+0.4}_{-0.05}$~M$_\odot$) with a luminosity of 45.7~L$_\odot$ \reply{at} a distance of 163~pc \citep{Gaia2018, Guzman-Diaz21}.  Its age is suggested to be $\sim$4~Myr \citep{Guzman-Diaz21}. \citet{Pacheo-Vazquez16} detected SO emission within the disk and subsequent observations by \citet{Riviere-Marichalar2020} show that SO emission is confined to within a ring corresponding to 100~au and extending to larger distances.  There is no CS measurement in this system and \citet{Riviere-Marichalar2020} \reply{find solutions that match current constraints with C/O=0.7 and  C/O=1.} We adopt C/O=0.7 for display in Fig.~\ref{fig:obsco}. The CO emission in the ring implies a gas-to-dust mass ratio of $\sim$40 \citep{Riviere-Marichalar2020}; this could be consistent with gas loss from the canonical factor of 100 in the interstellar medium or a small reduction in the CO gas phase abundance.
The models of \citet{Riviere-Marichalar2020} and \citet{Riviere-Marichalar2022} described above suggest a snowline location beyond 200~au in this source.

\subsection{TW Hya}

TW Hya is the closest young planet-forming disk system at a distance of 59.5 pc \citep{Gaia16}.  The spectral type is M0.5 (M = 0.6$\pm$0.1~M$_\odot$) with a luminosity of L = 0.26~L$\odot$ \citep{Sokal18, Herczeg14}.   The age of this system is debated \citep{Debes13, Herczeg14}; we list here the most recent estimate of 5-11 Myr \citep{Sokal18}.   The C/O ratio in TW Hya is estimated via a variety of means.  \citet{Bergin16} and \citet{Kama16_co} utilize \ce{C2H} emission to estimate 1.5-2.0 (we adopt 1.75).  \ce{C2H} emission in this system is found in a ring extending from $\sim 20-100$~au \citep{Kastner15, Bergin16}.  \citet{Cleeves21} presented detailed observations of \ce{C3H2} which traces comparable chemistry to \ce{C2H} and demonstrate that the elevated C/O ratio extends from 30~au to $\sim$120~au. \citet{Lee21} discuss how the $^{14}$N/$^{15}$N ratio observed in HCN has a dependence on the C/O ratio and suggest that C/O $>$ 1 extends into 20 au.  Interior to 20~au observations of water vapor and other tracers from Spitzer \citep{Carr11} which were analyzed by \citet{Bosman19_twhya}.  They find depleted 
carbon and oxygen in the inner 2.4~au with an overall solar C/O ratio.   We fix C/O to 1.75 into 20 au and \reply{to solar} inside 2.4 au.  This is also supported by analysis of abundances inside the silicate sublimation zone by \citet{McClure20}.  The connection between these two levels is uncertain.  The C/H abundance as traced by CO is of order 10$^{-6}$ interior and beyond the CO snowline \citep{Zhang19, Yoshida22}.  The snowline in this system is based on the $^{13}$C$^{18}$O emission and is suggested to lie near 21~au \citep{Zhang17}.

\subsection{LkCa 15}

LkCa 15 is a K5 star (M = 1.03~M$_\odot$) with a stellar luminosity of 1~L$_\odot$, an estimated age of 1-5 Myr  \citep{Simon2019, Donati2019, Pegues2020}, \reply{and is located at a distance of 159.2 $\pm$ 1.2~pc} \citep{Gaia2018}. This disk holds a large inner gap out to $\sim$50~au \citep{Pietu2006, Facchini2020}.   \citet{Sturm22} use \ce{C2H} to determine a C/O ratio of unity and a CO (C/H) abundance of 3.0$\pm1.5$ $\times$ 10$^{-5}$ from 50 au and beyond.  This CO abundance is at least a factor of 3 depleted relative to the general ISM CO abundance \citep[see][for discussion of ISM CO abundance]{Bergin17}.
\citet{Sturm22} and \citet{Qi19} estimate the CO snowline to lie near 58~au in this disk based on  analysis of CO and N$_2$H$^+$ emission.

\subsection{IM Lup}
IM Lup (Sz82) is a K5 star (M = 1.1~M$_\odot$) with a stellar luminosity of 2.57 $L_\odot$ at a distance of 158 pc \citep{Alcala17, Gaia2018, Oberg21}.  This system is generally estimated to have a young age of order 1-3 Myr \citep{Pinte08, Alcala17}.  \citet{Cleeves16_imlup}
 and \citet{Cleeves18} explored the overall physical structure and C/O ratio in this system, respectively.   The C/O ratio is generally estimated from \ce{C2H} but also is constrained via HCN observations, with an underpinning from CO isotopologue observations.   Most efforts to determine the C/O ratio from \ce{C2H} utilize a coarse grid ranging from 0.5, 1, to $> 1$.   In this effort \citet{Cleeves18} allowed for the presence of water ice in upper layers to provide oxygen (via photodesorption) to the system and suggesting that this provides needed dynamic range between 0.5 and 1.  The best fit was obtained between 40 to 100~au with C/O = 0.8.
   The inner tens of au of the IM Lup disk are obscured by strong dust continuum emission \citep{Cleeves16_imlup}.   The CO abundance (C/H) is estimated to be depleted by factors of 10-100 throughout the disk \citep{Cleeves16_imlup, Zhang21_mapsco}.   The CO snowline is found near 15~au based on analysis of CO isotopologue emission by \citet{Zhang21_mapsco}.

\subsection{AS 209}

AS 209 is a K5 star (M = 1.2~M$_\odot$; d = 121~pc) with a stellar luminosity of 1.4 L$_\odot$ and an estimated age of 1-3~Myr \citep{Gaia2018, Andrews18, Huang18, Avenhaus18, Oberg21}.  The C/O ratio was estimated via C$_2$H by \citet{Bosman21_mapsco} and \citet{Alarcon21}.  These works used independent codes but the same baseline physical model of the disk \citep[taken from][]{Zhang21_mapsco}.   Inside 10-20~au the drop in C$_2$H emission is interpreted as a drop in C/O towards 0.5.  Similarly, the C$_2$H emission exhibits a sharp drop in intensity beyond 100 au which can be modeled with decrease of C/O to 0.5.  \reply{Inside 20-100~au the C/O ratio is estimated to be $\sim$1.9.}  There is potential radial structure in the C/H or CO abundance. \citet{Zhang21_mapsco} and \citet{Alarcon21} suggest that the CO abundance is depleted by about a factor of 10 within the inner 100 au but rises to a factor a few depletion near 150~au and declines towards the outer disk.   The CO snowline is estimated to lie near 12~au based on analysis of CO isotopologue emission by \citet{Zhang21_mapsco}.

\subsection{Miotello C$_2$H Sample}
\citet{Miotello19} searched for \ce{C2H} emission towards a sample of disks with strong $^{13}$CO emission isolated in the Lupus survey by \citet{Ansdell16}.  In a total of nine targets \ce{C2H} emission was detected towards seven sources.  \citet{Miotello19} performed generic modeling of this sample to generate a relation between \ce{C2H}
 integrated flux density, disk mass, and overall C/O ratio.  The range of potential C/O ratios is shown in Fig.~\ref{fig:obsco}.  Two sources with non-detections could be interpreted as due to reduced C/O and we give that limit in the figure.

\section{Additional Landscape: Gaseous Disk C/H}

Another important aspect linking disk and exoplanet gas is the overall C/H ratios, for which there also are existing measurements.  Again, we focus on measurements obtained beyond the CO$_2$ snowline.  In this gas the expectation is that CO is the primary gas phase carbon carrier both inside and outside the CO snowline.   That is in layers where the dust temperature is above the sublimation temperature of CO we anticipate that the abundance of CO would be $\sim$50\% of the \reply{solar} value.  The remaining 50\% is found in refractory organic material \citep{Pollack94, Bergin15}.

A major complication in absolute abundance measurements is the estimation of the disk gaseous H$_2$ content.  H$_2$ is unemmissive in gas \reply{with} temperatures $\sim$20--50~K which encompasses the majority of the gas mass.   At present the best method to determine the H$_2$ mass  \citep[see discussion in][]{Bergin17, Miotello_ppvii} is via HD emission \citep{Bergin2013, McClure16, Trapman17}, pressure broadening of CO emission \citep{Yoshida22_pressure}, gas kinematics \citep{Paneque-Carreno21}, and N$_2$H$^{+}$ emission \citep{Trapman22, Anderson22}.

Based on these methods,  current suggestions are that the disk gas phase C/H ratio is depleted relative to the expectations set by the interstellar medium/stellar \citep{favre13a, McClure16, Schwarz16, Zhang17,
Yoshida22_pressure, Trapman22, Anderson22} by factors of a few to 100 \citep{Zhang19}. 
Of particular mention is the direct measurement of the H$_2$ density and (depleted) CO abundance in TW Hya inside of 20 au by \citet{Yoshida22_pressure}.
These are listed source by source in Appendix~C. The central question is whether the carbon carried by CO is found on small grains in the form of less volatile species \citep{Bergin14, Furuya14, Reboussin15, Eistrup17, Schwarz18, Bosman18} or is carried (as ice) by successively larger grains to the dust-rich midplane \citep{Xu17,Krijt18}.  This point matters in the context of the \citet{omb11} model as if CO were carried to the midplane by dust evolution (the latter solution) then it would be present in the solid planet cores.  However, if CO is found in less volatile form on small grains then it would be accreted alongside the gas during the phase of gas capture.    At present the situation is not clear.   It is certain that the dust mass is concentrated in the midplane \citep{Villenave20} and the ice must be as well.   Complex dust evolution models suggest dust evolution can matter and carry additional volatiles to the midplane \citep{Krijt20}, but other analyses suggested otherwise \citep{Ruaud22, Pascucci23}.
In our exploration we adopt C/H as a variable encompassing the range of highly depleted C/H (factor of 100) to relatively undepleted (factor of 2).  This factor of 2 encompasses the fact that 50\% of interstellar carbon is found in solid state carbonaceous grains \citep{Mishra15}.

\section{Exoplanet C/O Ratios}
\label{sec:exo_CO}
%We adopt recent measurements C/O and absolute abundances from a sample of direct detection planets and one Hot Jupiter (WASP-77Ab).   These are listed with references in Table~\ref{table:exo}. In cases where absolute abundances are not provided, they are instead estimated from the metallicity provided in the given reference. 

We adopt recent measurements C/O and absolute abundances from a sample of direct detection planets \reply{including the four HR~8799 planets, $\beta$~Pic~b, and HIP~65426~b}. These are listed with references in Table~\ref{table:exo}.
\reply{
For HR~8799 and $\beta$~Pic~b, the C/H abundances are derived from the metallicities, [M/H], given in the original references as the retrievals were done such that changes in the C/O changed the oxygen abundances only. Since \citet{Petrus21} do not state whether [C/H], [O/H], or both were modified when changing the C/O ratio, we have assumed that C/H is given by [M/H]. The assumption is not critical for our analysis, however.}

\reply{The abundances for the different planets have been determined from relatively homogeneous datasets. For HR~8799, we use the recent homogeneous analyses presented by \citet{Nasedkin2024} based on new GRAVITY data and supported by further data from SPHERE, GPI, CHARIS, OSIRIS, and ALES, and covering a wavelength range of 1--4$\micron$. The $\beta$~Pic~b study is similarly based on GRAVITY data, combined with GPI Y, J, \& H data and covers the range 1--2.5$\micron$, as described by \citet{Gravity20}. For HIP~65426~b, the data cover 1--4.7\micron, including SPHERE IFS data from 1--1.5\micron, SPHERE H band data, SINFONI K-band spectra and photometry in the L and M bands from NACO \citep{Petrus21}. Note the analysis of HIP~65426~b by \citet{Blunt2023} that also includes GRAVITY data arrives at similar results ($[{\rm M/H}] \approx 0.15$, $[{\rm C/O}] \approx 0.6$).
}

\reply{
The atmospheric compositions are derived from the spectra using a variety of different techniques: \citet{Petrus21} use fits to grids of Exo-REM \citep{Baudino2015} forward models to determine the abundances, while \citet{Gravity20} conduct both free retrievals with petitRADTRANS \citep{Molliere2019} and forward models with Exo-REM, finding consistent results. \citet{Nasedkin2024} explore different methodologies for determining the abundances of the HR~8799 planets, including free retrievals, chemical (dis-)equilibrium retrievals and fits to grids of forward models. Their results highlight an important issue: cloud condensation can significantly affect the atmospheric C/O ratio as oxygen is locked into silicates, complicating the interpretation of the inferred abundances. In particular, the retrievals favor C/O ratios 0f 0.7--0.8, while the forward models are better fit with C/O ratios nearer solar ($\approx 0.55$). 
}

\reply{
Where possible, we adopt the results from retrieval analyses because the fits derived from a given forward model presented by \citet{Nasedkin2024} are quoted with uncertainties that are much smaller than the systematic differences between the different models. As a result, we use the free retrieval for $\beta$~Pic-b, the dis-equilibrium chemistry retrieval for the HR~8799 system and the Exo-REM grid of forward models for  HIP~65426~b. For HR~8799, we therefore correct the oxygen abundances in our planet formation models to account for the condensation of silicate clouds. }

\begin{deluxetable}{rrrrcccccc}
\tablecolumns{10}
\tablewidth{0pt}
\tablecaption{Exoplanet Atmospheric Composition Measurements \label{table:exo}}
\tablehead{
\colhead{Name} & \colhead{a (au)} & \colhead{M$_p$ (M$_{Jup}$)} & \colhead{C/O} & \colhead{$\sigma$(C/O)} & \colhead{C/H\tablenotemark{a}} & \colhead{$\sigma$(C/H)} & \colhead{O/H\tablenotemark{a}} & \colhead{$\sigma$(O/H)} & \colhead{References\tablenotemark{b}}}
\startdata
HR8799e  & \reply{16.2} & \reply{7.5$^{+0.6}_{-0.6}$} &   \reply{0.83}  & \reply{0.02}    & \reply{125.89}  & \reply{73.63}  &  \reply{83.35}  & \reply{52.01} & (1)\\
HR8799d  & \reply{26.7} & \reply{9.2$^{+0.1}_{-0.1}$} & \reply{0.68}  & \reply{0.04}    & \reply{31.62} & \reply{18.50}  &   \reply{25.56}  & \reply{17.48}   &  (1)\\
HR8799c  & \reply{41.4} &\reply{8.5$^{+0.4}_{-0.4}$}  & \reply{0.62} &  \reply{0.01}    & \reply{11.22}  & \reply{1.08}  &   \reply{9.95}  & \reply{1.14} &    (1)\\
HR8799b  &  \reply{71.6} & \reply{6.0$^{+0.3}_{-0.3}$} &  \reply{0.73}  & \reply{0.02}    & \reply{14.45} & \reply{2.92}  &   \reply{10.88}  & \reply{2.57}  &  (1)\\
 $\beta$ Pic b & \reply{10.0}  & 9.3$^{+2.6}_{-2.5}$   &  0.43&   0.03    & 4.79 & 0.46 & {\it 6.12} & {\it 1.09} &   (2)\\
HIP65426b & \reply{92.0} &9.9$^{+1.1}_{-1.8}$ & $<$0.55& \nodata & 1.12 & 0.26 & \reply{$>$1.12} & \nodata & (3)\\
\enddata
\tablenotetext{a}{Abundances given as absolute relative to solar, i.e. (C/H)$_{\rm planet}$/(C/H)$_\odot$.  Adopted solar values are log$_{10}$(C/H) = $-$3.57 and log$_{10}$(O/H) = $-$3.31 from \citet{asplund09}. Values in italics were computed from the C/H and C/O quoted in the original references.}
\tablenotetext{b}{(1) \citet{Nasedkin2024}.  For masses adopt the single best fit retrieval parameters.   (2) \citet{Gravity20, Brandt2021_betapic}, (3) \citet{Petrus21, Wang2023_hip65426}}
\end{deluxetable}

\section{Planet Composition Model}

\reply{We adopt a simplified picture of planet formation where the bulk composition of the planets is computed by adding different amounts of gases and solids to the planet. We do not consider any radial variations in disc composition, so planet migration need not be considered. When comparing to the atmospheric composition we consider the impact of silicate cloud condensation on the observed abundances. }

\reply{We need the composition of the gases and solids in the disc to compute the planet compositions. Starting from the \cite{asplund09} solar abundances, $X_{\odot, i}$ (the number of atoms of species $i$ relative to the number of $H$ atoms), we partition the abundances into the gas phase, $X_{{\rm g}, i}$ and solid phase, $X_{{\rm s}, i}$. We assume that hydrogen and all noble gases are entirely in the gas phase (i.e. $X_{{\rm g}, i}=X_{\odot, i}$ when $i \in \{{\rm H, He, Ne, Xe, Ar}\}$). For nitrogen, 90\% is assumed to be in the gas-phase as $N_2$ (i.e. $X_{\rm g, N} = 0.9 X_{\odot, {\rm N}}$). The gas-phase carbon and oxygen abundances are determined by the  C/O ratio and the carbon depletion factor, $\Delta_{\rm C}$. Since approximately 50\% of carbon in the ISM is in a refractory form \citep{Mishra15}, we define $X_{\rm g, C} = X_{\odot, \rm C} / (2 \Delta_{\rm C})$. The gas-phase oxygen abundance is then determined by the C/O ratio, $X_{\rm g, O} = X_{\rm g, C} / (C/O)$. No other species are considered to be in the gas phase.}

\reply{The composition of the dust depends on whether the depletion of dust is considered. Without depletion, the total composition equates to solar so $X_{{\rm s}, i} = X_{\odot, i} - X_{{\rm g}, i}$.}

\reply{When computing the abundance of solids after dust depletion, we need to know how much material is lost together with its composition. We assume that any dust evolution occurs while the disc has its initial composition as this maximizes the impact of dust evolution. We assume the composition is inherited from the ISM and that most of the volatile carbon is in the form of CO. Defining the initial gas- and solid-phase composition as $X_{{\rm g}, i}^0$ and $X_{{\rm s}, i}^0$, we can compute the $X_{{\rm g}, i}^0$ by using $\Delta_{\rm C} = 1$ and ${\rm C/O}=1$ in the expressions for $X_{{\rm g}, i}$. Similarly, $X_{{\rm s}, i}^0 = X_{\odot, i} - X_{{\rm g}, i}^0$.}

\reply{We assume that dust depletion reduces the dust mass by a factor $\Delta_{\rm d}$. As a result, the total gas$+$solid composition after dust depletion is $X_{{\rm g}, i}^0 + X_{{\rm s}, i}^0 / \Delta_{\rm d}$. Dust enhancement through traps can be modelled via $\Delta_{\rm d} < 1.$ Next, the disc composition evolves to its current state with a given C/O and level of carbon depletion. At this point  $X_{{\rm s}, i}$ can be computed from $X_{{\rm g}, i}$ (itself calculated from $\Delta_{\rm C}$, C/O and $X_{\odot, \rm i}$, as above) and the new total disc abundance via $X_{{\rm g}, i} + X_{{\rm s}, i} = X_{{\rm g}, i}^0 + X_{{\rm s}, i}^0 / \Delta_{\rm d}$. Any further evolution of the dust-to-gas ratio in the disc does not affect the range of planet abundances compatible with the disc model and is thus neglected.}

\reply{The planet's bulk abundances, $X_{{\rm p}, i}$ are constructed by adding different amounts of gas and solids,  $X_{{\rm p}, i} = X_{{\rm g}, i} + f X_{{\rm s}, i}$, with the factor $f$ being varied to explore the different planet compositions compatible with the disc model. $f=1$ corresponds to the case where the disc and planet composition are the same. Silicate cloud condensation is accounted for using the prescription from \citet{Calamari2024}, in which condensation is estimated to reduce $X_{\rm p,O}$ by $2.024X_{\rm p,Si} + 1.167X_{\rm p,Mg}$, equivalent to 3.45 oxygen atoms per silicon atom.}

\reply{Finally, the mass fraction of solids accreted by the planet, $f_s$ is computed via 
\begin{equation}
f_s = \frac{\sum_i f X_{{\rm s}, i} m_i}{\sum_i m_i \left(X_{{\rm g}, i} + f X_{{\rm s}, i}\right)},
\end{equation}
where $m_i$ is mass of the species. Note that silicates are included in $X_{\rm p, O}$ when computing $f_{\rm s}$.}

\bibliography{zoverleaf}{}
\bibliographystyle{aasjournal}



\end{document}